\documentclass[12pt,journal,onecolumn]{IEEEtran}
\usepackage{epsfig,rotating,setspace,latexsym,amsmath,epsf,amssymb,amsfonts,bm,theorem,epstopdf}
\usepackage{cite,authblk}
\usepackage{algorithmic,algorithm}
\usepackage{setspace}
\usepackage{color}
\usepackage{graphicx}
\usepackage{mathrsfs}

\ifCLASSOPTIONcompsoc
\usepackage[caption=false, font=normalsize, labelfont=sf, textfont=sf]{subfig}
\else
\usepackage[caption=false, font=footnotesize]{subfig}
\fi

\newtheorem{theorem}{Theorem}
\newtheorem{lemma}{Lemma}

\newenvironment{Proof}[1]{\medskip\par\noindent{\bf Proof:\,}\,#1}{{\mbox{\,$\blacksquare$}\par}}

\allowdisplaybreaks

\begin{document}

\title{Selective Encoding Policies for Maximizing Information Freshness  \thanks{This work was supported by NSF Grants CCF 17-13977 and ECCS 18-07348, and is presented in part \cite{MelihBatu1} at Conference on Information Sciences and Systems, Princeton, NJ, March 2020. }}
\author{Melih Bastopcu \qquad Baturalp Buyukates  \qquad Sennur Ulukus\\
	\normalsize Department of Electrical and Computer Engineering\\
	\normalsize University of Maryland, College Park, MD 20742\\
	\normalsize  \emph{bastopcu@umd.edu}\hspace{0.5em} \qquad \emph{baturalp@umd.edu}  \qquad \hspace{0.5em} \emph{ulukus@umd.edu}}
\maketitle

\begin{abstract}	
 An information source generates independent and identically distributed status update messages from an observed random phenomenon which takes $n$ distinct values based on a given pmf. These update packets are encoded at the transmitter node to be sent to a receiver node which wants to track the observed random variable with as little age as possible. The transmitter node implements a selective $k$ encoding policy such that rather than encoding all possible $n$ realizations, the transmitter node encodes the most probable $k$ realizations. We consider three different policies regarding the remaining $n-k$ less probable realizations: \emph{highest $k$ selective encoding} which disregards whenever a realization from the remaining $n-k$ values occurs; \emph{randomized selective encoding} which encodes and sends the remaining $n-k$ realizations with a certain probability to further inform the receiver node at the expense of longer codewords for the selected $k$ realizations; and \emph{highest $k$ selective encoding with an empty symbol} which sends a designated empty symbol when one of the remaining $n-k$ realizations occurs. For all of these three encoding schemes, we find the average age and determine the age-optimal real codeword lengths, including the codeword length for the empty symbol in the case of the latter scheme, such that the average age at the receiver node is minimized. Through numerical evaluations for arbitrary pmfs, we show that these selective encoding policies result in a lower average age than encoding every realization, and find the corresponding age-optimal $k$ values. 
 
\end{abstract}
 
\section{Introduction}
Age of information is a performance metric which quantifies the timeliness of information in networks. It keeps track of the time since the most recent update at the receiver has been generated at the transmitter. Age increases linearly in time such that at time $t$ age $\Delta(t)$ of an update packet which was generated at time $u(t)$ is $\Delta(t) = t-u(t)$. When a new update packet is received, the age drops to a smaller value. Age of information has been studied in the context of queueing networks \cite{Kaul12a, Costa14, Bedewy16, He16a, Kam16b, Sun17a, Najm18b, Najm17, Soysal18, Soysal19}, scheduling and optimization \cite{Yates17b, Tang19, Nath17, Hsu18b, Kadota18a, Buyukates18c,Buyukates19b, Gong19, Buyukates19c, Arafa19b, Sun17b, Sun18b, Chakravorty18, Bastopcu19, Bastopcu20b, Bastopcu20c, partial_updates, Zou19b, Non_linear, Bastopcu18,  bastopcu_soft_updates_journal}, energy harvesting \cite{Arafa17b, Arafa17a, Wu18, Arafa_Age_Online, Arafa18a, Arafa18f, Arafa19e, Farazi18, Yener_energy_19, Chen19}, reinforcement learning \cite{Elmagid18, Liu18, Ceran18, Beytur19, Elmagid19} problems and so on. The concept of age is applicable to a wide range of problems, e.g., in autonomous driving, augmented reality, social networks, and online gaming, as information freshness is crucial in all these and other emerging applications.

\begin{figure}[t]
	\centering  \includegraphics[width=0.7\columnwidth]{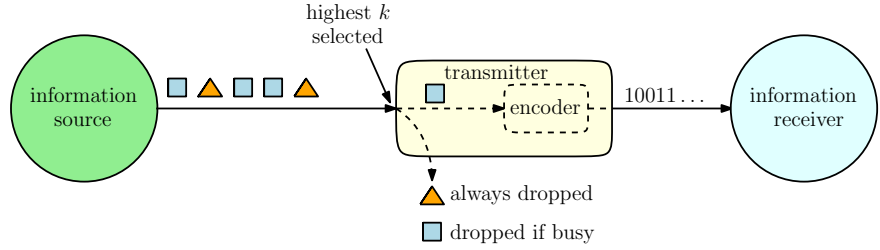}
	\caption{An information source generates i.i.d. status updates from a random variable $X$. Only a portion of the realizations (shown with a square) is encoded into codewords. Update packets that come from the selected portion of the realizations that find the transmitter node idle are sent to the receiver node. Non-selected realizations (shown with a triangle) are always discarded at the transmitter node even if the transmitter node is idle. }
	\label{fig:model}
\end{figure}

In this work, we consider a status updating system that consists of a single transmitter node and a single receiver node (see Fig.~\ref{fig:model}). The transmitter receives independent and identically distributed time-sensitive status update packets generated by an information source based on an observed random phenomenon that takes $n$ distinct values with a known pmf. This observed random variable could be the position of a UAV in autonomous systems or share prices in the stock market.  Arriving status update packets are encoded at the transmitter and sent to the receiver through an error-free noiseless channel. The receiver wants to acquire fresh information regarding the observed random variable, which brings up the concept of age of information.

Unlike most of the literature in which the transmission times, also referred to as service times in queueing theory, are based on a given service distribution, in this work, we design transmission times through source coding schemes by choosing the codeword lengths assigned to realizations. That is, the codeword length assigned to each realization represents the service time (transmission time) of that realization. 

References that are most closely related to our work are \cite{Mayekar18, Zhong16, Yates_Soljanin_source_coding} which study the timely source coding problem. Reference \cite{Mayekar18} considers zero-wait update generation and finds codeword lengths using Shannon codes based on a modified version of the given pmf that achieve the optimal age with a constant gap. References \cite{Zhong16} and \cite{Yates_Soljanin_source_coding} consider block coding and source coding problems to find age-optimal codes for FIFO queues. 

Inspired by these works, we first introduce a \emph{selective encoding} mechanism at the transmitter node. In this selective encoding model, instead of encoding all possible realizations, we encode only a portion of the realizations and send to the receiver node. Specifically, we consider what we call the \emph{highest $k$ selective encoding} scheme in which we only encode the most probable $k$ realizations and disregard any update packets from the remaining $n-k$ realizations. Similar $k$ out of $n$ type of schemes are shown to achieve good age performance especially in the context of multicast networks in which each update packet is transmitted until the earliest $k$ of the $n$ receiver nodes receive that packet \cite{Zhong17a, Zhong18b, Buyukates18, Buyukates18b, Buyukates19}. We note that a smaller $k$ in our setting yields shorter codeword lengths but larger interarrival times, as in this case most of the updates are not encoded. However, when $k$ is large, codeword lengths and correspondingly the transmission times get larger even though the interarrival times get smaller. Thus, in this paper, based on the given pmf, we aim to find the optimal $k$ which strikes a balance between these two opposing trends such that the average age at the receiver node is minimized. Due to this selective encoding scheme not every realization is sent to the receiver even if the channel is free.

Next, we consider a scenario in which the remaining $n-k$ realizations are not completely disregarded but encoded with a certain probability which we call the \emph{randomized selective encoding} scheme. In this scheme, in addition to the most probable $k$ realizations, the remaining $n-k$ less probable realizations are sometimes encoded. 

\begin{figure}[t]
	\centering  \includegraphics[width=0.7\columnwidth]{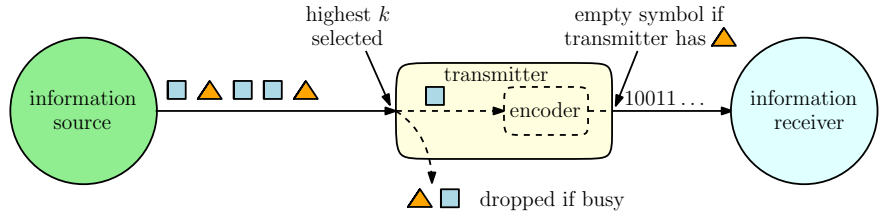}
	\caption{Update packets that come from the selected portion of the realizations (shown with a square) that find the transmitter idle are sent to the receiver. Non-selected realizations (shown with a triangle) that find the transmitter idle are mapped into an empty symbol.}
	\label{fig:model_emptysymbol}
\end{figure}

A disadvantage of the highest $k$ selective encoding scheme is the fact that the receiver node is not informed when one of the non-selected realizations occurs. For instance, during a period with no arrivals, the receiver node cannot differentiate whether there has been no arrivals or if the arrival has taken one of the non-selected values as in either case it does not receive any update packets. Thus, lastly, we take a careful look at the remaining $n-k$ realizations and propose a modified selective encoding policy which we call the \emph{highest $k$ selective encoding with empty symbol} that still achieves a lower average age than encoding every realization but also informs the receiver node when one of the non-selected values is taken by the observed random variable. In this scheme, only the most probable $k$ realizations are encoded and the remaining $n-k$ realizations are mapped into a designated empty symbol such that in the case of these $n-k$ non-selected realizations, this empty symbol is sent to further inform the receiver (see Fig.~\ref{fig:model_emptysymbol}). Thus, in such a case, the receiver at least knows that the observed random variable has taken a value from the non-selected portion even though it does not know which value was taken specifically. We consider two variations on this policy: when the empty symbol does not reset the age and when the empty symbol resets the age.

For all three encoding schemes, we find the average age experienced by the receiver node and determine the age-optimal real codeword lengths, including the codeword length of the empty symbol in the case of the highest $k$ selective encoding with empty symbol scheme. Through numerical evaluations for given arbitrary pmfs, we show that the proposed selective encoding policies achieve a lower average age than encoding every realization, and find the corresponding age-optimal $k$ values.        
\section{System Model and Problem Formulation}\label{sect:model}
We consider a communication system in which an information source generates independent and identically distributed status update packets from an observed phenomenon that takes realizations from the set $\mathcal{X}= \{x_1, x_2, \ldots, x_n\}$ based on a known pmf $P_X(x_i)$ for $i \in \{1, \ldots, n\}$. Without loss of generality, we assume that $P_X(x_m) \geq P_X(x_j)$ for all $m\leq j$, i.e., the probabilities of the realizations are in a non-increasing order. Update packets arrive at the transmitter node following a Poisson process with parameter $\lambda$. The transmitter node implements a blocking policy in which the update packets that arrive when the transmitter node is busy are blocked and lost. Thus, the transmitter node receives only the updates which arrive when it is idle. 

We consider three different encoding policies: highest $k$ selective encoding, randomized selective encoding, and highest $k$ selective encoding with an empty symbol.
\subsection{Policy 1: Highest $k$ Selective Encoding}\label{Subsect:1}
In the first policy, we consider a selective encoding mechanism that we call \emph{highest k selective encoding} where the transmitter node only sends the most probable $k$ realizations, i.e., only the realizations from set $\mathcal{X}_k =\{x_1, \ldots, x_k\}$, which have the highest probabilities among possible $n$ updates generated by the source, are transmitted for $k \in \{1,\ldots, n\}$; see Fig.~\ref{fig:model}. If an update packet from the remaining non-selected portion of the realizations arrives, the transmitter disregards that update packet and waits for the next arrival. If an update packet arrives from the selected portion of the realizations, then the transmitter encodes that update packet by using a binary alphabet with the conditional probabilities given by, 
\begin{align}\label{cond_prob}
   P_{X_k}(x_i)= \begin{cases} 
      \frac{P_X(x_i)}{q_k}, & i=1,2,\ldots, k \\
      0, & i=k+1,k+2,\ldots, n,   
   \end{cases}
\end{align}
where 
\begin{align}\label{eqn:q_k}
q_k \triangleq \sum_{\ell=1}^{k}P_X(x_\ell).    
\end{align}
The transmitter assigns codeword $c(x_i)$ with length $\ell(x_i)$ to realization $x_i$ for $i\in\{1,2,\ldots,k\}$.
\subsection{Policy 2: Randomized Selective Encoding}\label{Subsect:2}
In the second policy, inspired by \cite{Mayekar18}, we study a \emph{randomized selective encoding} scheme. In this policy, the most probable $k$ realizations are always encoded. However, instead of discarding the remaining $n-k$ realizations, the transmitter node encodes them with probability $\alpha$ and discards them with probability $1-\alpha$. In other words, in this model, less likely realizations that are not encoded under the highest $k$ selective encoding policy are sometimes transmitted to the receiver node. Thus, under this operation, codewords for each one of the $n$ possible realizations need to be generated since every realization can be sent to the receiver node. The transmitter assigns codeword $c(x_i)$ with length $\ell(x_i)$ to realization $x_i$ for $i\in\{1,2,\ldots,n\}$.

The transmitter node performs encoding using the following conditional probabilities,
\begin{align}\label{cond_prob2}
   P_{X_\alpha}(x_i)= \begin{cases} 
      \frac{P_X(x_i)}{q_{k,\alpha}}, & i=1,2,\ldots, k \\
      \alpha \frac{P_X(x_i)}{q_{k,\alpha}}, & i=k+1,k+2,\ldots, n,   
   \end{cases}
\end{align}
where
\begin{align}
 q_{k,\alpha} \triangleq \sum_{\ell=1}^{k}P_X(x_\ell) +\alpha \sum_{\ell=k+1}^{n} P_X({x_\ell}).   
\end{align}

\subsection{Policy 3: Highest $k$ Selective Encoding with an Empty Symbol}\label{Subsect:3}
In the third policy, we consider an encoding scheme that we call the \emph{highest k selective encoding with an empty symbol}. In this encoding scheme, the transmitter always encodes the most probable $k$ realizations as in the previous two policies. However, unlike the previous models, if an update packet from the remaining non-selected portion of the realizations arrives, the transmitter sends an empty status update denoted by $x_e$ to further inform the receiver at the expense of longer codewords for the selected $k$ realizations. 

When an update packet arrives from the set $ \mathcal{X}_k' = \mathcal{X}_k \cup \{x_e\}$, the transmitter node encodes that update packet with the binary alphabet by using the pmf given as $\{P_X(x_1),P_X(x_2),\ldots,P_X(x_k), \\ P_X(x_e)\}$ where $P_X(x_e)=1-q_k$. Thus, in this policy, the transmitter node assigns codewords to the most probable $k$ realizations as well as to the empty symbol $x_e$. That is, the transmitter assigns codeword $c(x_i)$ with length $\ell(x_i)$ to realization $x_i$ for $i\in\{1,\ldots,k,e\}$.

In this paper, we focus on the source coding aspect of timely status updating. Therefore, in all these three policies, the channel between the transmitter node and the receiver node is error-free. The transmitter node sends one bit at a unit time. Thus, if the transmitter node sends update $x_i$ to the receiver node, this transmission takes $\ell(x_i)$ units of time. That is, for realization $x_i$, the service time of the system is $\ell(x_i)$. 

\subsection{Problem Formulation}
We use the age of information metric to measure the freshness of the information at the receiver node. Let $\Delta(t)$ be the instantaneous age at the receiver node at time $t$ with $\Delta(0)= \Delta_0$. Age at the receiver node increases linearly in time and drops to the age of the most recently received update upon delivery of a new update packet. We define the long term average age as,
\begin{align}
    \Delta = \lim_{T\to\infty} \frac{1}{T}\int_0^T\Delta(t)dt.
\end{align}

Our aim is to find the codeword lengths for each encoding policy described in Sections \ref{Subsect:1}, \ref{Subsect:2}, and \ref{Subsect:3} that minimize the long term average age for a given $k$ such that a uniquely decodable code can be designed, i.e., the Kraft inequality is satisfied \cite{Cover}. 

In the following section, we find an analytical expression for the long term average age $\Delta$. 

\section{Average Age Analysis}\label{sect:age_analysis}
As described in Section~\ref{sect:model}, status update packets arrive at the transmitter as a Poisson process with rate $\lambda$. Update packets that arrive when the transmitter is busy are blocked from entry and dropped. Thus, upon successful delivery of a packet to the receiver, the transmitter idles until the next update packet arrives. This idle waiting period in between two arrivals is denoted by $Z$ which is an exponential random variable with rate $\lambda$ due to the memoryless property of exponential random variables as update interarrivals at the transmitter are exponential with $\lambda$. 

We note that in all of the encoding policies in Section \ref{sect:model}, every packet from the set $\mathcal{X}_k$ which successfully enters the transmitter node is always sent to the receiver. However, a packet from the remaining least probable $n-k$ realizations which enters the transmitter might not be sent. Under the highest $k$ selective encoding policy described in Section \ref{Subsect:1}, when one of the remaining $n-k$ packets enters the transmitter node, the transmitter node drops the packet and proceeds to wait for the next update arrival. Under the randomized selective encoding scheme described in Section \ref{Subsect:2}, remaining $n-k$ less likely realizations are transmitted to the receiver node with probability $\alpha$. Under the highest $k$ selective encoding scheme with an empty symbol described in Section \ref{Subsect:3}, the transmitter node sends a designated empty status update to further inform the receiver about the occurrence of a realization from the remaining $n-k$ realizations.  

\begin{figure}[t]
	\centering  \includegraphics[width=0.5\columnwidth]{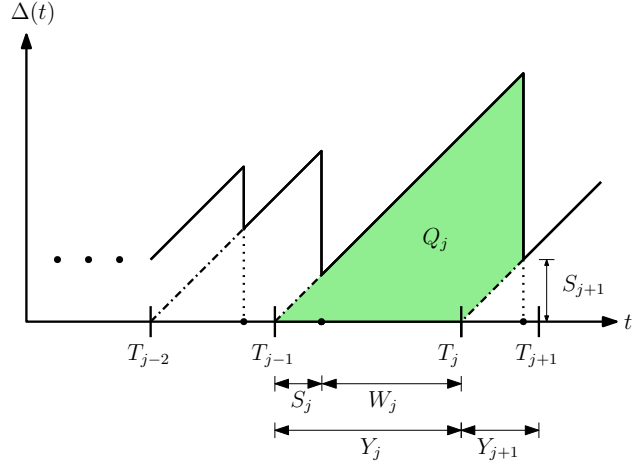}
	\caption{Sample age evolution $\Delta{(t)}$ at the receiver node. Successful updates are indexed by $j$. The $j$th successful update arrives at the server node at $T_{j-1}$. Update cycle at the server node is the time in between two successive arrivals and is equal to $Y_j = S_j + W_j = T_{j} - T_{j-1}$.}
	\label{fig:ageEvol}
\end{figure}

We denote the update packets which arrive when the transmitter node is idle and reset the age as successful update packets. Since the channel is noiseless and there is no preemption, these successful packets are received by the receiver node. We denote $T_{j-1}$ as the time instant at which the $j$th successful update packet is received. We define update cycle denoted by $Y_j = T_j - T_{j-1}$ as the time in between two successive successful update arrivals at the transmitter. Update cycle $Y_j$ consists of a busy cycle and an idle cycle such that
\begin{align}
	Y_j = S_j + W_j, \label{update_cycle}
\end{align}
where $S_j$ is the service time of update $j$ and $W_j$ is the overall waiting time in the $j$th update cycle.

Fig.~\ref{fig:ageEvol} shows a sample age evolution at the receiver. Here, $Q_j$ denotes the area under the instantaneous age curve in update cycle $j$ and $Y_j$ denotes the length of the $j$th update cycle as defined earlier. The metric we use, long term average age, is the average area under the age curve which is given by \cite{Najm17}
\begin{align}
	\Delta = \limsup_{n\to\infty} \frac{\frac{1}{n}\sum_{j=1}^{n} Q_j}{\frac{1}{n}\sum_{j=1}^{n}Y_j} = \frac{\mathbb{E}[Q]}{\mathbb{E}[Y]}. \label{avg_age1}
\end{align}
By using Fig.~\ref{fig:ageEvol}, we find $ Q_j = \frac{1}{2}Y^2_j + Y_j S_{j+1}$, where $Y_j$ is given in (\ref{update_cycle}). Thus, using the independence of $Y_j$ and $S_{j+1}$, (\ref{avg_age1}) is equivalent to
\begin{align}
    \Delta =  \frac{\mathbb{E}[Y^2]}{2\mathbb{E}[Y]}+\mathbb{E}[S]. \label{avg_age2}
\end{align}

In the following section, we find the optimal real-valued codeword lengths for the highest $k$ selective encoding policy described in Section \ref{Subsect:1}.
\section{Optimal Codeword Design Under Selective Encoding} \label{sect:opt_soln}
In this section, we consider the highest $k$ selective encoding policy described in Section \ref{Subsect:1}. Under this way of operation, the transmitter only sends the most probable $k$ realizations from the set $\mathcal{X}_k$, and drops any update packets from the remaining $n-k$ least probable realizations.  

Theorem \ref{thm1} characterizes the average age $\Delta$ given in (\ref{avg_age2}) for the encoding scheme described in Section~\ref{Subsect:1}.
\begin{theorem}\label{thm1}
    Under the highest $k$ selective encoding scheme, the average age at the receiver node is given by
    \begin{align}
        \Delta =  \frac{\mathbb{E}[L^2] + \frac{2}{q_k\lambda} \mathbb{E}[L] + \frac{2}{(q_k\lambda)^2}}{2\left(\mathbb{E}[L] +\frac{1}{q_k\lambda}\right)}+\mathbb{E}[L], \label{avg_age3} 
    \end{align}
 where the first and the second moments of the codeword lengths are given by $\mathbb{E}[L] =  \sum_{i=1}^{k}P_{X_k}(x_i) \\ \ell(x_i)$ and $\mathbb{E}[L^2] = \sum_{i=1}^{k}P_{X_k}(x_i)\ell(x_i)^2$.
\end{theorem}{}
\begin{Proof}
With the highest $k$ selective encoding scheme, we note that the overall waiting time $W$ is eqaul to $W = \sum_{\ell=1}^{M}Z_{\ell}$. Here, $M$ is a geometric random variable with parameter $q_k$ (defined in (\ref{eqn:q_k})) which denotes the total number of 
update arrivals until the first update from the set $\mathcal{X}_k$ is observed at the transmitter node. $W$ is also an exponential random variable with rate $\lambda q_k$ \cite[Prob. 9.4.1]{Yates14}. The moments of $M$ are
\begin{align}
\mathbb{E}[M] &= \frac{1}{{q_k}}, \label{moment1} \\  
\mathbb{E}[M^2] &= \frac{2-{q_k}}{q_k^2}. \label{moment2}
\end{align}
We note that the first and second moments of the service time are equal to the corresponding moments of the codeword lengths, i.e., $\mathbb{E}[S] = \mathbb{E}[L]$ and $\mathbb{E}[S^2] = \mathbb{E}[L^2]$, as the transmitter is capable of sending one bit at a unit time. By inspecting Fig.~\ref{fig:ageEvol} and noting that $M$, $L$ and $Z$ are mutually independent, we find \cite[Thm. 9.11]{Yates14}
\begin{align}
      \mathbb{E}[Y] &= \mathbb{E}[L] + \mathbb{E}[M]\mathbb{E}[Z], \label{E[Y]}\\
      \mathbb{E}[Y^2] &= \mathbb{E}[L^2] + 2\mathbb{E}[M]\mathbb{E}[Z]\mathbb{E}[L]+\mathbb{E}[M]\mathbb{E}[Z^2] +(\mathbb{E}[M^2]-\mathbb{E}[M])(\mathbb{E}[Z])^2\label{E[Y^2]},
\end{align}
where moments of $M$ follow from (\ref{moment1}) and (\ref{moment2}), and $Z$ has exponential distribution with rate $\lambda$ as discussed earlier. Substituting (\ref{E[Y]}) and (\ref{E[Y^2]}) in (\ref{avg_age2}) yields the result in (\ref{avg_age3}).
\end{Proof}{}

Thus, (\ref{avg_age3}) characterizes the average age $\Delta$ achieved at the receiver node in terms of the first and second moments of the codeword lengths for a given pmf, selected $k$, and update arrival rate $\lambda$. Next, we formulate the age minimization problem as,
\begin{align}
\label{problem1_mod}
\min_{\{ \ell(x_i) \}}  \quad &  \frac{\mathbb{E}[L^2]+2a\mathbb{E}[L]+2a^2}{2(\mathbb{E}[L]+a)} +\mathbb{E}[L] \nonumber \\
\mbox{s.t.} \quad & \sum_{i=1}^{k} 2^{-\ell(x_i)}\leq 1 \nonumber \\
\quad & \ell(x_i) \in \mathbb{R}^+, \quad i\in\{ 1,\ldots, k\},
\end{align}
where the objective function is equal to the average age found in Theorem \ref{thm1} with $a = \frac{1}{\lambda q_k}$, the first constraint is the Kraft inequality, and the second consraint represents the feasibility of the codeword lengths, i.e., each codeword length should be non-negative.

Similar to \cite{Sun17b} and \cite{Arafa_Age_Online}, we define $p(\theta)$ as
\begin{align}\label{p_theta}
    p(\theta):=& \min_{\{ \ell(x_i) \}}\frac{1}{2}\mathbb{E}[L^2]+\mathbb{E}[L]^2+(2a-\theta)\mathbb{E}[L]+a^2-\theta a \nonumber\\
    &\quad \mbox{s.t.} \quad \sum_{i=1}^{k} 2^{-\ell(x_i)}\leq 1 \nonumber \\
    & \qquad \quad \hspace{0.5em} \ell(x_i) \in \mathbb{R}^+, \quad i\in\{ 1,\ldots, k\}.
\end{align}
One can show that $p(\theta) $ is decreasing in $\theta$ and the optimal solution is obtained when $p(\theta) = 0$ such that the optimal age for the problem in (\ref{problem1_mod}) is equal to $\theta$, i.e., $\Delta^*= \theta $ \cite{frac_programming}. We define the Lagrangian \cite{Boyd04} function as
\begin{align}
    \mathcal{L} =&\frac{1}{2}\mathbb{E}[L^2]+\mathbb{E}[L]^2+(2a-\theta)\mathbb{E}[L]+a^2-\theta a +\beta\left(\sum_{i=1}^{k} 2^{-\ell(x_i)}- 1  \right),
\end{align}
where $\beta \geq 0$. Next, we write the KKT conditions as
\begin{align}\label{KKT_cond}
    \frac{\partial\mathcal{L}}{\partial \ell(x_i)} =&  P_{X_k}(x_i)\ell(x_i)+2\mathbb{E}[L]P_{X_k}(x_i) + (2a-\theta)P_{X_k}(x_i)-\beta (\log2)2^{-\ell(x_i)} = 0, \quad \forall i. 
\end{align}
The complementary slackness condition is
\begin{align}\label{CS_cond}
   \beta\left(\sum_{i=1}^{k} 2^{-\ell(x_i)}- 1  \right) = 0. 
\end{align}

In the following lemma, we prove that the optimal codeword lengths must satisfy the Kraft inequality as an equality. 
\begin{lemma}\label{Lemma:CS}
	For the age-optimal real codeword lengths, we must have $ \sum_{i=1}^{k} 2^{-\ell(x_i)}= 1$.
\end{lemma} 
\begin{Proof}
	Assume that the optimal codeword lengths satisfy $\sum_{i=1}^{k} 2^{-\ell(x_i)}<1$, which implies that $\beta= 0$ due to (\ref{CS_cond}). From (\ref{KKT_cond}), we have
	\begin{align}
	P_{X_k}(x_i)\ell(x_i)+2\left(\sum_{j=1}^{k}P_{X_k}(x_j)\ell(x_j)\right)P_{X_k}(x_i)
    + (2a-\theta)P_{X_k}(x_i) = 0, \quad \forall i.
	\end{align}
	Then, we find $\ell(x_i)= \frac{\theta-2a}{3}$ for all $i \in \{1,2,\ldots,k\}$. Thus, $\mathbb{E}[L]= \frac{\theta-2a}{3}$ and $ \mathbb{E}[L^2]=\left( \frac{\theta-2a}{3}\right)^2$ so that $p(\theta) = -\frac{\theta^2}{6}-\frac{\theta a}{3} + \frac{a^2}{3}$. By using $p(\theta) = 0$,  we find  $\theta = (-1+\sqrt{3})a$ which gives $\ell(x_i)=\frac{(-3+\sqrt{3})a}{3} < 0$ for $i\in\{1,2,\ldots, k\} $. Since the codeword lengths cannot be negative, we reach a contradiction. Thus, the optimal codeword lengths must satisfy $\sum_{i=1}^{k} 2^{-\ell(x_i)}=1$.   
\end{Proof}

Next, we find the optimal codeword lengths which satisfy $\sum_{i=1}^{k} 2^{-\ell(x_i)}= 1$. By summing (\ref{KKT_cond}) over all $i$, we obtain
\begin{align}\label{E_L_thr}
  \mathbb{E}[L] = \frac{\theta +\beta \log2-2a}{3}.  
\end{align}
From (\ref{KKT_cond}), we obtain
 \begin{align} \label{KKT_mod}
     -\ell(x_i) +\frac{\beta\log2}{P_{X_k}(x_i)} 2^{-\ell(x_i)} = 2\mathbb{E}[L]+2a-\theta,
 \end{align}
 for $i\in\{1,2,\ldots, k\}$, which yields
 \begin{align}\label{KKT_mod2}
     \frac{\beta (\log2)^2}{P_{X_k}(x_i)}2^{-\ell(x_i)}e^{\frac{\beta (\log2)^2}{P_{X_k}(x_i)}2^{-\ell(x_i)}} = \frac{\beta (\log2)^2}{P_{X_k}(x_i)}2^{\frac{-\theta+2\beta\log2+2a}{3}}. 
 \end{align}
Note that (\ref{KKT_mod2}) is in the form of $xe^x=y$ where the solution for $x$ is equal to $x = W_0(y)$ if $y\geq 0$. Here, $W_0(\cdot)$ denotes the principle branch of the Lambert $W$ function \cite{lambert}. Since the right hand side of (\ref{KKT_mod2}) is always non-negative, we are only interested in $W_0(\cdot)$ which is denoted as $W(\cdot)$ from now on. We find the unique solution for $\ell(x_i)$ as
\begin{align}\label{eqn:opt_lengths}
\ell(x_i) =-\frac{\log \left( \frac{ P_{X_k}(x_i)}{\beta (\log2)^2} W\left(\frac{\beta (\log2)^2}{P_{X_k}(x_i)} 2^{\frac{-\theta +2\beta \log2+2a}{3}} \right)\right)}{\log2},
\end{align} 
for $i\in\{1,2,\ldots, k\}$. 

In order to find the optimal codeword lengths, we solve (\ref{eqn:opt_lengths}) for a $(\theta, \beta)$ pair that satisfies $p(\theta)  = 0$ and the Kraft inequality, i.e., $\sum_{i=1}^{k} 2^{-\ell(x_i)} = 1$. Starting from an arbitrary $(\theta, \beta)$ pair, if $p(\theta)>0$ (or $p(\theta)<0$), we increase (or respectively decrease) $\theta$ in the next iteration as $p(\theta)$ is a decreasing function of $\theta$. Then, we update $\beta$ by using (\ref{E_L_thr}). We repeat this process until $p(\theta)  = 0$ and $\sum_{i=1}^{k} 2^{-\ell(x_i)} = 1$.   

We note that the age-optimal codeword lengths found in this section are for a fixed $k$. Thus, depending on the selected $k$, different age performances are achieved at the receiver node. In Section~\ref{sect:num_res}, we find the age-optimal $k$ values for some given arbitrary pmfs numerically.

Under the highest $k$ selective encoding policy, the receiver node does not receive any update when the remaining $n-k$ realizations occur. However, there may be scenarios in which these remaining realizations are also of interest to the receiver node. In the next section, we focus on this scenario and consider a randomized selection of the remaining $n-k$ realizations so that these realizations are not completely ignored.

\section{Optimal Codeword Design under Randomized Selective Encoding}\label{sect:randomized}
The selective encoding scheme discussed so far is a deterministic scheme in which a fixed number of realizations are encoded into codewords and sent to the receiver node when realized. In this section, inspired by \cite{Mayekar18}, we consider a randomized selective encoding scheme where the transmitter encodes the most probable $k$ realizations with probability $1$, and encodes the remaining least probable $n-k$ realizations with probability $\alpha$ and thus, neglects them with probability $1-\alpha$. Thus, this randomized selective encoding policy strikes a balance between encoding every single realization and the highest $k$ selective encoding scheme discussed so far.  

Theorem \ref{thm_random} determines the average age experienced by the receiver node under the randomized highest $k$ selective encoding scheme.

\begin{theorem}\label{thm_random}
    Under the randomized highest $k$ selective encoding scheme, the average age at the receiver node is given by
    \begin{align}
        \Delta_\alpha =  \frac{\mathbb{E}[L^2] + \frac{2}{q_{k,\alpha}\lambda} \mathbb{E}[L] + \frac{2}{(q_{k,\alpha}\lambda)^2}}{2\left(\mathbb{E}[L] +\frac{1}{q_{k,\alpha}\lambda}\right)}+\mathbb{E}[L], \label{avg_age5} 
    \end{align}
where $\mathbb{E}[L] = \sum_{i=1}^n P_{X_\alpha}(x_i) \ell(x_i)$, and  $\mathbb{E}[L^2] = \sum_{i=1}^n P_{X_\alpha}(x_i) \ell(x_i)^2$.
\end{theorem}{}
The proof of Theorem~\ref{thm_random} follows similarly to that of Theorem~\ref{thm1} by replacing $q_{k}$ with $q_{k,\alpha}$. 

Next, we formulate the age minimization problem for this case as,
\begin{align}
\label{problem3_mod}
\min_{\{ \ell(x_i), \alpha \}}  \quad &  \frac{\mathbb{E}[L^2]+2\bar{a}\mathbb{E}[L]+2\bar{a}^2}{2(\mathbb{E}[L]+\bar{a})} +\mathbb{E}[L] \nonumber \\
\mbox{s.t.} \quad & \sum_{i=1}^{n} 2^{-\ell(x_i)}\leq 1 \nonumber \\
\quad & \ell(x_i) \in \mathbb{R}^+, \quad i\in\{ 1,\ldots, n\},
\end{align}
where the objective function is equal to the average age $\Delta_\alpha$ in Theorem \ref{thm_random} with $\bar{a} = \frac{1}{\lambda q_{k,\alpha}}$, the first and second constraints follow from the Kraft inequality and the feasibility of the codeword lengths, i.e., each codeword length should be non-negative.

We first solve this problem for a fixed $\alpha$ in this section and determine the optimal $\alpha$ numerically for given arbitrary pmfs in Section~\ref{sect:num_res}. Following a similar solution technique to that in Section~\ref{sect:opt_soln}, we find
\begin{align}\label{eqn:opt_lengths3}
\ell(x_i) =-\frac{\log \left( \frac{ P_{X_\alpha}(x_i)}{\beta (\log2)^2} W\left(\frac{\beta (\log2)^2}{P_{X_\alpha}(x_i)} 2^{\frac{-\theta +2\beta \log2+2\bar{a}}{3}} \right)\right)}{\log2},
\end{align} 
for $i\in \{1,2,\ldots, n\}$. To determine the age-optimal codeword lengths $\ell(x_i)$ for $i \in \{1, 2, \ldots, n\}$, we then employ the algorithm described in Section~\ref{sect:opt_soln}.

In the following section, we consider the case where instead of sending the remaining least probable $n-k$ realizations randomly, the transmitter sends an empty symbol for these updates to further inform the receiver. 

\section{Optimal Codeword Design Under Selective Encoding with an Empty Symbol} \label{sect:empty_status}
In this section, we calculate the average age by considering two different scenarios for the empty symbol. Operationally, the receiver may not reset its age when $x_e$ is received as it is not a regular update packet and the receiver does not know which realization occurred specifically. On the other hand, the receiver may choose to update its age as this empty symbol carries some information, the fact that the current realization is not one of the $k$ encoded realizations, regarding the observed random variable. Thus, in this section, we consider both of these scenarios\footnote{We note that another possible scenario may be to drop the age to an intermediate level between not updating at all and updating fully considering the partial information conveyed by the empty status update. This case is not considered in this paper.} 
and find the age-optimal codeword lengths for the set $ \mathcal{X}_k'$ with the pmf $\{P_X(x_1),P_X(x_2),\ldots,P_X(x_k), P_X(x_e)\}$ in each scenario.
\subsection{When the Empty Symbol Does Not Reset the Age}\label{subsect:doesnot_reset}
In this way of operation, the age at the receiver is not updated when the empty status update $x_e$ is received. Thus, sending $x_e$ incurs an additional burden since it does not reset the age but increases the average codeword length of the selected $k$ realizations.

The update cycle is given by (\ref{update_cycle}) with
\begin{align} \label{eqn:waiting_time}
    W = (M-1)\ell(x_e)+\sum_{\ell=1}^{M}Z_\ell,
\end{align} 
where $M$ is defined in Section \ref{sect:opt_soln} and denotes the total number of update arrivals until the first update from the set $\mathcal{X}_k$ is observed at the transmitter. In other words, there are $M-1$ deliveries of the empty status update $x_e$ in between two successive deliveries from the encoded set $\mathcal{X}_k$. As discussed earlier, $Z$ is an exponential random variable with rate $\lambda$ and $M$ is a geometric random variable with parameter $q_k$. By using the fact that the arrival and service processes are independent, i.e., $S$ and $Z$ are independent, and $M$ is independent of $S$ and $Z$, in Theorem~\ref{thm3}, we find the average age when an empty status update does not reset the age.
\begin{theorem}\label{thm3}
    When the empty status update $x_e$ does not reset the age, the average age under the highest $k$ selective encoding scheme with an empty symbol at the receiver is given by
    \begin{align}
        \Delta_e =  \frac{\mathbb{E}[L^2|X_k' \neq x_e] + 2\mathbb{E}[W] \mathbb{E}[L|X_k' \neq x_e] +\mathbb{E}[W^2]} {2\left(\mathbb{E}[L|X_k' \neq x_e] + \mathbb{E}[W]\right)}+\mathbb{E}[L|X_k' \neq x_e] . \label{avg_age4} 
    \end{align}
\end{theorem}{}
\begin{Proof}
We note that the service time of a successful update is equal to its codeword length so that we have
\begin{align}
\mathbb{E}[S] =& \mathbb{E}[L|X_k' \neq x_e] = \sum_{i=1}^{k}P_{X_k}(x_i)\ell(x_i) \label{service_1}\\
\mathbb{E}[S^2] =& \mathbb{E}[L^2|X_k' \neq x_e]=\sum_{i=1}^{k}P_{X_k}(x_i)\ell(x_i)^2 \label{service_2}
\end{align}
 where $ P_{X_k}(x_i)$ is defined in (\ref{cond_prob}). By using the independence of $M$ and $Z$, we find
\begin{align}
      \mathbb{E}[W] =& \ell(x_e)\left(\frac{1}{q_k}-1\right)+\frac{1}{\lambda
       q_k}, \label{E[W]}\\
      \mathbb{E}[W^2] =& \frac{(2-q_k)(1-q_k)}{q_k^2}\ell(x_e)^2+\frac{4(1-q_k)}{\lambda q_k^2}\ell(x_e)+\frac{2}{(\lambda q_k)^2}\label{E[W^2]},
\end{align}
where the moments of $M$ follow from (\ref{moment1}) and (\ref{moment2}), and $Z$ has exponential distribution with rate $\lambda$. Substituting (\ref{service_1})-(\ref{E[W^2]}) in (\ref{avg_age2}) yields the result in (\ref{avg_age4}).
\end{Proof}{}

We note that $\Delta_e$ in (\ref{avg_age4}) depends on $\ell(x_e)$ only through the overall waiting time $W$ as the age does not change when $x_e$ is received. Next, we write the age minimization problem as
\begin{align}
\label{problem2}
\min_{\{ \ell(x_i), \ell(x_e) \}}  \quad & \frac{\mathbb{E}[L^2|X_k' \neq x_e] + 2\mathbb{E}[W] \mathbb{E}[L|X_k' \neq x_e] +\mathbb{E}[W^2]} {2\left(\mathbb{E}[L|X_k' \neq x_e] + \mathbb{E}[W]\right)} +\mathbb{E}[L|X_k' \neq x_e] \nonumber \\
\mbox{s.t.} \quad & 2^{-\ell(x_e)} + \sum_{i=1}^{k} 2^{-\ell(x_i)}\leq 1 \nonumber \\
\quad & \ell(x_i)\in \mathbb{R}^+, \quad i\in\{1,\ldots,k,e \},
\end{align}
where the objective function is equal to the average age expression $\Delta_e$ in (\ref{avg_age4}). We note that problem (\ref{problem2}) is not convex due to the middle term in the objective function. However, when $\ell(x_e)$ is fixed, it is a convex problem. Thus, we first solve the problem in (\ref{problem2}) for a fixed $\ell(x_e)$ and then determine the optimal $\ell(x_e)$ numerically in Section~\ref{sect:num_res}. 

Thus, for a fixed $\ell(x_e)$, (\ref{problem2}) becomes
\begin{align}
\label{problem2_mod}
\min_{\{ \ell(x_i) \}}  \quad & \frac{\mathbb{E}[L^2|X_k' \neq x_e] + 2\mathbb{E}[W] \mathbb{E}[L|X_k' \neq x_e] +\mathbb{E}[W^2]} {2\left(\mathbb{E}[L|X_k' \neq x_e] + \mathbb{E}[W]\right)} +\mathbb{E}[L|X_k' \neq x_e] \nonumber \\
\mbox{s.t.} \quad & \sum_{i=1}^{k} 2^{-\ell(x_i)}\leq 1-2^{-c} \nonumber \\
\quad & \ell(x_i) \in \mathbb{R}^+, \quad i\in\{1,\ldots,k\},
\end{align}
where $\ell(x_e) = c$. Since the empty status update length $\ell(x_e)$ is fixed and given, we write the Kraft inequality by subtracting the portion allocated for $\ell(x_e)$ in the optimization problem in (\ref{problem2_mod}). Similar to previous sections, we define $p(\theta)$ as 
\begin{align}\label{p_kappa}
    p(\theta):=& \min_{\{ \ell(x_i) \}}\frac{1}{2}\mathbb{E}[L^2|X_k' \neq x_e] +\mathbb{E}[L|X_k' \neq x_e]^2 +(2\hat{a}-\theta)\mathbb{E}[L|X_k' \neq x_e]+\frac{d}{2}-\theta \hat{a}   \nonumber\\
    &\quad \mbox{s.t.} \quad \sum_{i=1}^{k} 2^{-\ell(x_i)}\leq 1-2^{-c} \nonumber \\
    & \qquad \quad \hspace{0.5em} \ell(x_i) \in \mathbb{R}^+,\quad i\in\{1,\ldots,k\},
\end{align}
where $\hat{a}= \mathbb{E}[W]$ and $d = \mathbb{E}[W^2]$. For a fixed and given $\ell(x_e)$, the optimization problem in (\ref{p_kappa}) is convex. We define the Lagrangian function as
\begin{align}
    \mathcal{L} =& \frac{1}{2}\mathbb{E}[L^2|X_k' \neq x_e] +\mathbb{E}[L|X_k' \neq x_e]^2 +(2\hat{a}-\theta)\mathbb{E}[L|X_k' \neq x_e]+\frac{d}{2}-\theta \hat{a}   \nonumber\\ &+ \beta \left(\sum_{i=1}^{k} 2^{-\ell(x_i)}+2^{-c}-1 \right),
\end{align}
where $\beta \geq 0$. The KKT conditions are
\begin{align}\label{KKT_cond2}
    \frac{\partial\mathcal{L}}{\partial \ell(x_i)} =& P_{X_k}(x_i)\ell(x_i)+2\mathbb{E}[L|X_k' \neq x_e]P_{X_k}(x_i) + (2\hat{a}-\theta)P_{X_k}(x_i)-\beta (\log2)2^{-\ell(x_i)} = 0, 
\end{align}
for all $i$, and the complementary slackness condition is 
\begin{align}\label{CS_cond2}
    \beta\left( \sum_{i=1}^{k} 2^{-\ell(x_i)}+2^{-c}-1\right)=0.
\end{align}

Lemma~\ref{Lemma:CS2} shows that the optimal codeword lengths satisfy $\sum_{i=1}^{k} 2^{-\ell(x_i)}= 1-2^{-c}$.
\begin{lemma}\label{Lemma:CS2}
	For the age-optimal real-valued codeword lengths, we must have $ \sum_{i=1}^{k} 2^{-\ell(x_i)}= 1-2^{-c}$.
\end{lemma} 
\begin{Proof}
	Assume that the optimal codeword lengths satisfy $\sum_{i=1}^{k} 2^{-\ell(x_i)}<1-2^{-c}$, which implies that $\beta= 0$ due to (\ref{CS_cond2}). From (\ref{KKT_cond2}), we have
	\begin{align}\label{lemma1_eqn}
	P_{X_k}(x_i)\ell(x_i)+2\left(\sum_{j=1}^{k}P_{X_k}(x_j)\ell(x_j)\right)P_{X_k}(x_i)
    + (2\hat{a}-\theta)P_{X_k}(x_i) = 0, \quad \forall i.
	\end{align}
	By summing (\ref{lemma1_eqn}) over all $i$, we get $\mathbb{E}[L]= \frac{\theta-2\hat{a}}{3}$. Then, we find $\ell(x_i)= \frac{\theta-2\hat{a}}{3}$ for all $i \in \{1,\ldots,k\}$ which makes $p(\theta) = -\frac{\theta^2+2\hat{a}\theta+4\hat{a}^2-3d}{6}$. By using $p(\theta) = 0$,  we find  $\theta = -\hat{a}+\sqrt{3(d-\hat{a}^2)}$ which gives $\ell(x_i)=-\hat{a}+\sqrt{\frac{d-\hat{a}^2}{3}}$ for $i\in\{1,\ldots, k\} $. One can show that $\theta$, hence age, is a decreasing function of $c$. Thus, in the optimal policy, $c$ must be equal to zero. However, choosing $c=0$ leads to $\sum_{i=1}^{k}2^{-\ell(x_i)}< 1-2^{-c}=0$. Since the sum on the left cannot be negative, we reach a contradiction. Thus, the optimal codeword lengths must satisfy $\sum^k_{i=1} 2^{-\ell(x_i)}=1-2^{-c}.$ 
\end{Proof}

Thus, for the age-optimal codeword lengths, we have $ \sum_{i=1}^{k} 2^{-\ell(x_i)}= 1-2^{-c}$ and $\beta \geq 0$ from (\ref{CS_cond2}). By summing (\ref{KKT_cond2}) over all $i$ and using Lemma~\ref{Lemma:CS2} we find
\begin{align}\label{expect}
    \mathbb{E}[L|X_k' \neq x_e] = \frac{\theta+\beta \log2(1-2^{-c})-2\hat{a}}{3}.
\end{align}
From (\ref{KKT_cond2}), we obtain
 \begin{align} \label{KKT_mod_v2}
     -\ell(x_i) +\frac{\beta\log2}{P_{X_k}(x_i)} 2^{-\ell(x_i)} = 2\mathbb{E}[L|X_k' \neq x_e] +2\hat{a}-\theta.
 \end{align}
Thus, we find the unique solution for $\ell(x_i)$ as
\begin{align}\label{eqn:opt_lengths2}
    \ell(x_i) = -\frac{\log \left( \frac{P_{X_k}(x_i)}{\beta (\log2)^2}W\left(\frac{\beta (\log2)^2}{P_{X_k}(x_i)} 2^{\frac{-\theta +2\beta \log2(1-2^{-c})+2\hat{a}}{3}} \right)\right)}{\log 2} ,
\end{align}
for $i\in\{1,\ldots, k\}$. To determine the age-optimal codeword lengths $\ell(x_i)$ for $i \in \{1, \ldots, k\}$, we then employ the algorithm described in Section~\ref{sect:opt_soln}.

We note that the average age achieved at the receiver depends on $\ell(x_e)$. In Section~\ref{sect:num_res}, we provide numerical results where we vary $\ell(x_e)$ over all possible values and choose the one that yields the least average age for given arbitrary pmfs.
\subsection{When the Empty Symbol Resets the Age}
In this subsection, we consider the case where the empty symbol resets the age as it carries \textit{partial} status information as in \cite{partial_updates, Bastopcu20a}. In other words, each update which arrives when the transmitter idles is accepted as a successful update. 

Theorem~\ref{thm4} determines the average age $\Delta_e$ when the empty symbol resets the age. 

\begin{theorem}\label{thm4}
When the empty status update $x_e$ resets the age, the average age under the highest $k$ selective encoding scheme at the receiver is given by
    \begin{align}
        \Delta_e =&  \frac{\mathbb{E}[L^2]+2\frac{1}{\lambda}\mathbb{E}[L]+\frac{2}{\lambda^2}}{2\left(\mathbb{E}[L]+\frac{1}{\lambda}\right)} +\mathbb{E}[L]. \label{avg_age6} 
    \end{align}
\end{theorem}{}
\begin{Proof}
Different from the previous sections, the moments for the waiting time are equal to $\mathbb{E}[W] = \frac{1}{\lambda}$ and $\mathbb{E}[W^2] = \frac{2}{\lambda^2}$ as each successful symbol is able to reset the age. Thus, substituting $\mathbb{E}[W]$ and $\mathbb{E}[W^2]$ in (\ref{avg_age2}) and noting that $\mathbb{E}[S] = \mathbb{E}[L]$ yields the result.
\end{Proof}{}

Next, we formulate the age minimization problem as 
\begin{align}
\label{problem3}
\min_{\{ \ell(x_i),\ell(x_e) \}}  \quad &  \frac{\mathbb{E}[L^2]+2\tilde{a}\mathbb{E}[L]+2\tilde{a}^2}{2(\mathbb{E}[L]+\tilde{a})} +\mathbb{E}[L] \nonumber \\
\mbox{s.t.} \quad & 2^{-\ell(x_e)}+\sum_{i=1}^{k} 2^{-\ell(x_i)}\leq 1 \nonumber \\
\quad & \ell(x_i)\in \mathbb{R}^+,\quad i\in\{1,\ldots, k, e\},
\end{align}
where $\tilde{a} = \frac{1}{\lambda}$. We follow a similar solution technique to that given in Section \ref{sect:opt_soln} to get
\begin{align}\label{eqn:opt_lengths4}
\ell(x_i) =-\frac{\log \left( \frac{ P_{X}(x_i)}{\beta (\log2)^2}W\left(\frac{\beta (\log2)^2}{P_{X}(x_i)} 2^{\frac{-\theta +2\beta \log2+2\tilde{a}}{3}} \right)\right)}{\log2},
\end{align} 
for $i\in\{1,\ldots, k, e\}$.

The value of $k$ affects $\ell(x_e)$ such that when $k$ is close to $n$, the probability of the empty symbol becomes small which leads to a longer $\ell(x_e)$, whereas when $k$ is small, the probability of the empty symbol becomes large which results in a shorter $\ell(x_e)$. In Section~\ref{sect:num_res}, we numerically determine the optimal $k$ selection which achieves the lowest average age for a given arbitrary distribution.

\section{Numerical Results} \label{sect:num_res}

\begin{figure}[t]
	\centering  \includegraphics[width=0.7\columnwidth]{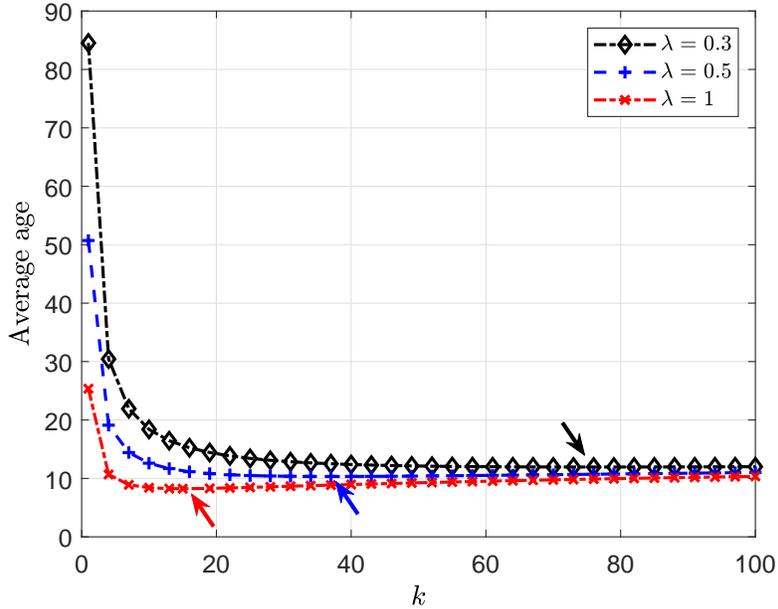}
	\caption{The average age values with the age-optimal codeword lengths for $\lambda \in \{0.3,0.5,1\}  $ for the pmf provided in (\ref{zipf_pmf}) with the parameters $n=100$, $s=0.4$. We vary $k$ from $1$ to $n$ and indicate $k$ that minimizes the average age for each $\lambda$ with an arrow.  }
	\label{sim1}
\end{figure}

In this section, we provide numerical results for the optimal encoding policies that are discussed in Sections \ref{sect:opt_soln}, \ref{sect:randomized}, and \ref{sect:empty_status}. In the first two numerical results, we perform simulations to characterize optimal $k$ values that minimize the average age with the highest $k$ selective encoding scheme in Section \ref{sect:opt_soln}. For these simulations, we use Zipf$(n,s)$ distribution with the following pmf for $n=100$, $s= 0.4$,
\begin{align}\label{zipf_pmf}
    P_X(x_i) = \frac{i^{-s}}{\sum_{j=1}^{n}j^{-s} }, \quad 1\leq i\leq N.
\end{align}

\begin{figure}[t]
	\centering  \includegraphics[width=0.7\columnwidth]{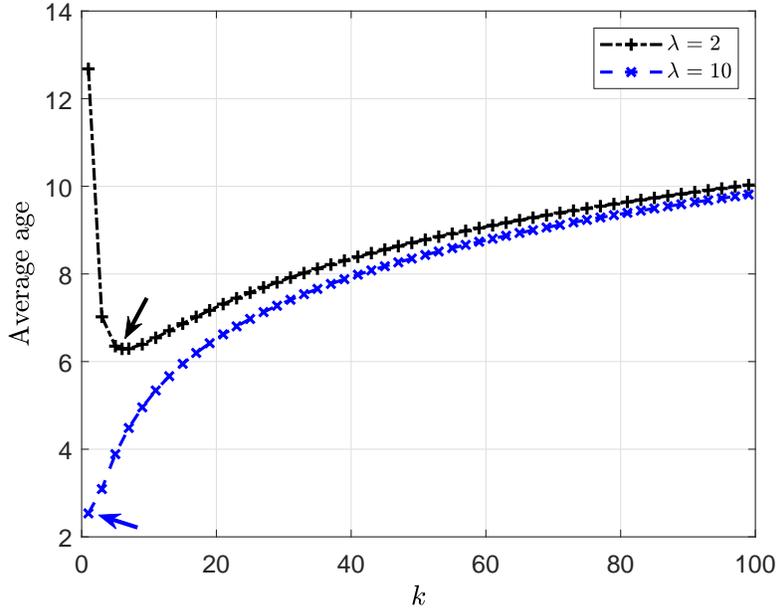}
	\caption{The average age values with the age-optimal codeword lengths for $\lambda \in \{2,10\}  $ for the pmf provided in (\ref{zipf_pmf}) with the parameters $n=100$, $s=0.4$. We vary $k$ from $1$ to $n$ and observe that choosing $k=1$ under the relatively high arrival rates ($\lambda =10$) minimizes the average age.}
	\label{sim2}
\end{figure}

In Fig.~\ref{sim1}, we show the effect of sending the most probable $k$ realizations when the update packets arrive at the transmitter node rather infrequently, i.e., the arrival rate is low. We consider the cases in which the arrival rate is equal to $\lambda = 0.3, 0.5,1$. For each arrival rate, we plot the average age as a function of $k=1,2,\ldots, n$. We see that increasing the arrival rate reduces the average age as expected. In this case, optimal $k$ is not equal to $1$ since the effective arrival rate is small. In other words, the transmitter node wants to encode more updates as opposed to idly waiting for the next update arrival when the arrivals are rather infrequent. Choosing $k$ close to $n$ is also not optimal as the service times of the status updates with low probabilities are longer which hurts the overall age performance. Indeed, in Fig.~\ref{sim1}, where update arrival rates are relatively small, it is optimal to choose $k=76$ for $\lambda = 0.3$, $k=37$ for $\lambda = 0.5$, and $k=15$ for $\lambda = 1$.        

\begin{figure}[t]
	\centering  \includegraphics[width=0.7\columnwidth]{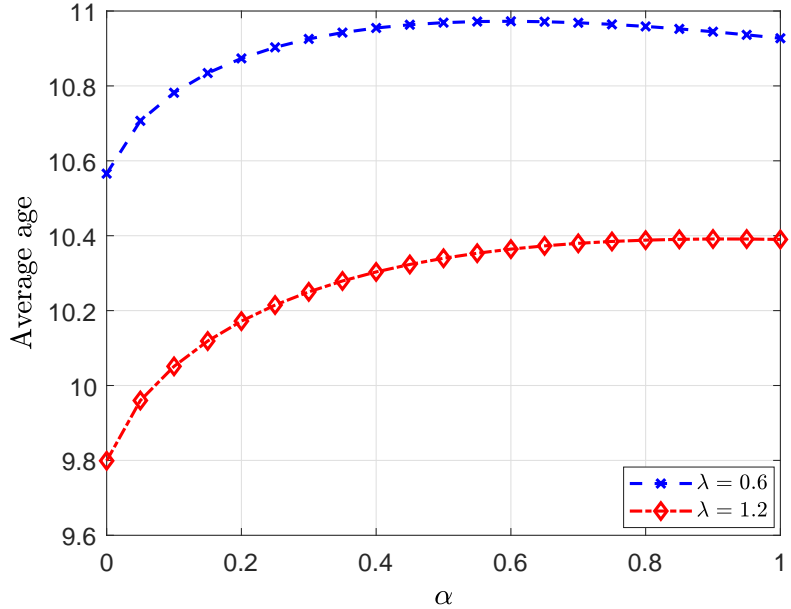}
	\caption{The average age values with the age-optimal codeword lengths for different $\alpha$ values with the pmf provided in (\ref{zipf_pmf}) with $n=100$, $s=0.2$ for $k=70$ when randomized highest $k$ selective encoding is implemented.}
	\label{sim5_randomized}
\end{figure}

In Fig.~\ref{sim2}, we consider a similar setting as in Fig.~\ref{sim1} but here update arrival rates are larger which means that updates arrive more frequently at the transmitter node. We observe that when $\lambda = 2$, the optimal $k$ is still not equal to $1$ (it is equal to 6 in Fig.~\ref{sim2}) as the updates are not frequent enough. However, once updates become more available to the transmitter node, i.e., the case with $\lambda=10$ in Fig.~\ref{sim2}, we observe that the transmitter node chooses to only encode the realization with the highest probability, i.e., $k=1$, and wait for the next update arrival instead of encoding more and more realizations which increases the overall codeword lengths thereby increasing the transmission times. We also observe that the average age decreases as the update arrival rate increases as in Fig.~\ref{sim1}.

For the third numerical result shown in Fig.~\ref{sim5_randomized}, we simulate the randomized highest $k$ selective encoding policy described in Section \ref{sect:randomized} with Zipf distribution in (\ref{zipf_pmf}) with parameters $n=100$, $s=0.2$. In Fig.~\ref{sim5_randomized}, we observe two different trends depending on the update arrival frequency at the source node, even though in either case, randomization results in a higher age at the receiver node than selective encoding, i.e., $\alpha=0$ case. When the arrival rate is high, $\lambda =1.2$ in Fig.~\ref{sim5_randomized}, we observe that age monotonically increases with $\alpha$ as randomization increases average codeword length, i.e., service times. Although increasing $\alpha$ results in a higher age at the receiver node, previously discarded $n-k$ realizations can be received under this randomized selective encoding policy. Interestingly, when the arrival rate is smaller, $\lambda =0.6$ in Fig.~\ref{sim5_randomized}, we observe that age initially increases with $\alpha$ and then starts to decrease because of the decreasing waiting times as opposed to increasing codeword lengths such that when $\alpha$ is larger than $0.3$, it is better to select $\alpha=1$, i.e., encoding every realization. That is, when $\alpha$ grows beyond $0.3$, encoding and sending every single realization yields a lower average age.

\begin{figure}[t]
	\centering  \includegraphics[width=0.7\columnwidth]{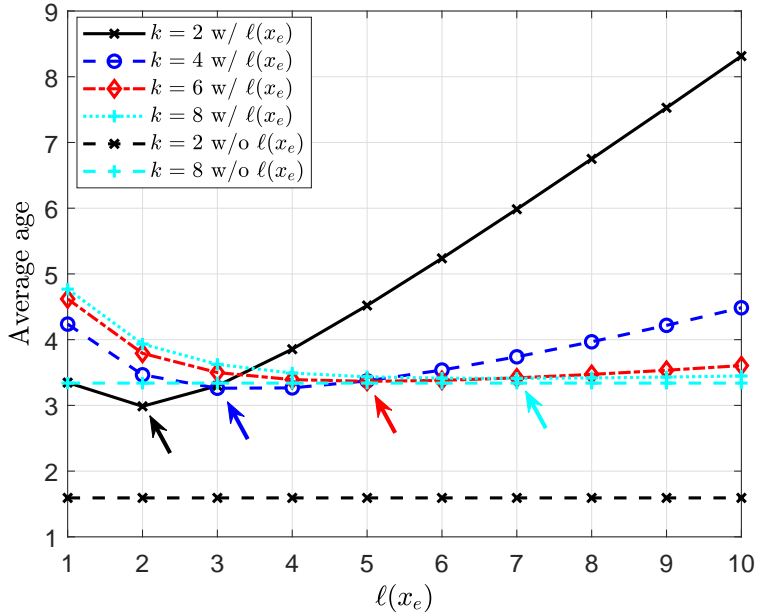}
	\caption{Average age with the age-optimal codeword lengths with respect to $\ell(x_e)$ with the pmf in (\ref{sim_pmf}) for $n=10$ when the empty symbol does not reset the age. Arrows indicate the age-optimal $\ell(x_e)$ values. We also provide the optimal age without sending the empty symbol for $k=2$ and $k=8$.}
	\label{sim3}
\end{figure}

In the fourth and fifth numerical results, we find the optimal real-valued codeword lengths and $k$ values that minimize the average age $\Delta_e$ with the highest $k$ selective encoding scheme with an empty symbol discussed in Section \ref{sect:empty_status}. For these numerical results, we use the following pmf
\begin{align}\label{sim_pmf}
    P_X(x_i) =\begin{cases} 
      2^{-i}, & i=1,\ldots, n-1 \\
      2^{-n+1}, & i=n.   
   \end{cases} 
\end{align}

In the fourth numerical result, we consider the pmf in (\ref{sim_pmf}) for $n=10$ and take $\lambda = 5$. We find the optimal codeword length of the empty symbol, $\ell(x_e)$, when the empty symbol does not reset the age (see Fig.~\ref{sim3}). We observe that when $k$ is small, the probability of sending the empty symbol becomes large so that a shorter codeword is preferable for $x_e$. For example, we observe in Fig.~\ref{sim3} that choosing $\ell(x_e) = 2$ when $k=2$ and $\ell(x_e) = 3$ when $k=4$ is optimal. Similarly, when $k$ is larger, a longer codeword is desirable for $x_e$. We observe in Fig.~\ref{sim3} that choosing $\ell(x_e) = 5$ when $k=6$ and $\ell(x_e) = 7$ when $k=8$ is optimal. Further, we note in Fig.~\ref{sim3} that the average age increases when we send the empty symbol in the case of the remaining $n-k$ realizations as the empty symbol increases the total waiting time for the next successful arrival as well as the codeword lengths for the encoded $k$ realizations. For smaller $k$ values, i.e., when $k=2$, this effect is significant as the empty symbol has a large probability whereas when $k$ is larger, i.e., when $k=8$, sending an empty status update increases the age slightly (especially when $\ell(x_e)$ is high) as the empty symbol has a small probability.

\begin{figure}[t]
	\centering  \includegraphics[width=0.7\columnwidth]{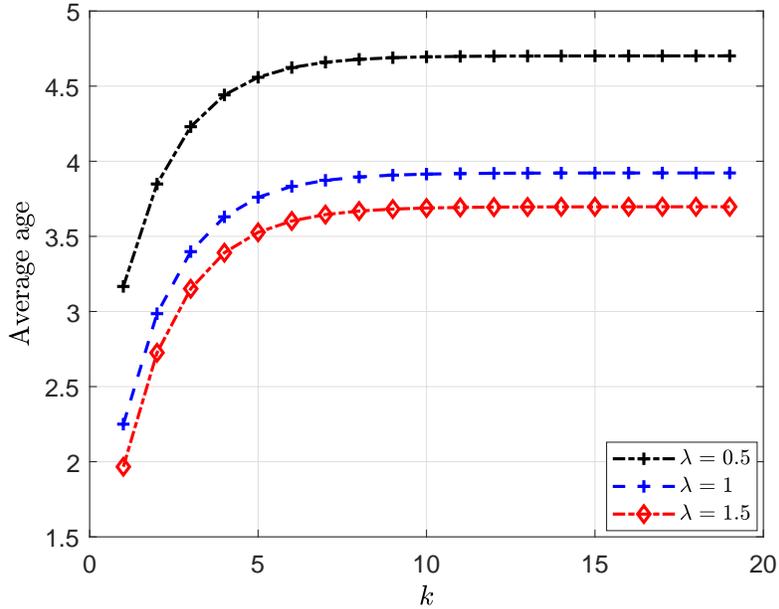}
	\caption{Average age with the age-optimal codeword lengths for varying $k$ with the pmf in (\ref{sim_pmf}) for $n=20$ when the empty symbol resets the age.}
	\label{sim4_empty_symbol_resets}
\end{figure}

In the fifth numerical result shown in Fig.~\ref{sim4_empty_symbol_resets}, we consider the case when the empty symbol $x_e$ resets the age. We observe that the minimum age is achieved when 
$k=1$, i.e., only the most probable realization is encoded. This is because the overall waiting time is independent of $k$ and larger $k$ values result in larger codewords which in turn increases transmission times. Thus, in this case, only the most probable realization is received separately since all others are embedded into the empty symbol. We note that this selection results in significant information loss at the receiver which is not captured by the age metric alone. This problem can be addressed by introducing a distortion constraint which measures the information loss together with the age metric which measures freshness \cite{Bastopcu20a}.

\section{On The Optimality of the Highest $k$ Selective Encoding}\label{sect:8}

So far, we have considered only the case where the most probable $k$ realizations are encoded and sent through the channel. Based on this selection, we found the average age and determined the age-optimal $k$ and codeword lengths. We observed that this highest $k$ selective encoding policy results in a lower average age than encoding every realization. However, we note that there are ${n \choose k}$ selections for encoding and in this section, we discuss the optimality of the highest $k$ selective encoding among all these different selections. We see that the average age expression in Theorem~\ref{thm1} depends on the pmf of $X$ which affects the optimal codeword lengths, and the effective arrival rate. In this section, we denote the effective arrival rate as $\lambda_e$ given by $\lambda_e = \lambda \sum_{x\in \mathcal{X}_s}P_X(x)$ where $\mathcal{X}_s$ is the set of arbitrarily selected $k$ updates for encoding. Here, by choosing a different set of $k$ realizations to encode and send, instead of the most probable $k$ realizations, we change the effective arrival rate and codeword lengths which in turn yields a different age performance.

When the arrival rate is relatively low, we see in Fig.~\ref{sim1} that the average age is dominated mainly by the effective arrival rate. Thus, choosing the realizations with the highest probabilities may be desirable as this selection achieves the highest possible effective arrival rate. However, when the arrival rate is relatively high, the average age is mainly determined by the moments of the codeword lengths. 

In Table~\ref{table1}, we find the age-optimal update selections for given pmfs and arrival rates for $k=5$. We use the pmf in (\ref{sim_pmf}) with $n=10$ and Zipf distribution in (\ref{zipf_pmf}) with parameters $n=10$, $s=0.2$. In both pmfs, the updates are in decreasing order with respect to their probabilities, i.e., $P_X(x_i) \geq P_X(x_j)$ if $i\leq j$. When the arrival rate is relatively small, i.e., $\lambda = 0.1$ for the first pmf and $\lambda = 0.5$ for the second pmf, we observe that choosing the realizations with the highest probabilities for encoding is optimal. That is, the optimal selection is $\{1, 2, 3, 4, 5 \}$ when $\lambda = 0.1$ for the first pmf and when $\lambda = 0.5$ for the second pmf. However, when the arrival rate is high, the optimal policy is to encode the realization with the highest probability and $k-1$ realizations with the lowest probabilities such that the optimal set is $\{1, 7, 8, 9, 10 \}$. We see that this selection is optimal when $\lambda = 1$ for the first pmf and when $\lambda =2$ for the second pmf. From these, we deduce that the optimal update selection strategy is to keep the effective arrival rate as high as possible while maintaining the moments of the codeword lengths at the desired levels. We see this structure when $\lambda = 0.5$ for the first pmf and $\lambda = 1$ for the second pmf where the optimal selection is choosing the most probable two and the least probable three realizations i.e., the optimal selection is $\{1, 2, 8, 9, 10 \}$.

\begin{table*}[t]
\centering
\begin{tabular}{ |c|c|c|c|c|c| } 
\hline
pmf & $\lambda$ & optimal selection & $\lambda_{e}$ & optimal age\\
\hline
The pmf in (\ref{sim_pmf}) for $n=10$ & $0.1$ & $\{1,2,3,4,5\}$ & $0.0969$ & $12.292$  \\ 
& $0.5$ & $\{1,2,8,9,10\}$ & $0.3789$ & $3.867$ \\ 
& $1$ & $\{1,7,8,9,10\}$ & $0.5156$ & $2.4229$  \\ 
\hline
Zipf($n=10$, $s=0.2$)& $0.5$ & $\{1,2,3,4,5\}$ & $0.3898$ & $5.154$  \\ 
& $1$ & $\{1,2,8,9,10\}$ & $0.6269$ & $3.929$ \\ 
& $2$ & $\{1,7,8,9,10\}$ & $1.01$ & $3.304$  \\  
\hline 
\end{tabular}
\vspace{1 mm}
  \caption{ \label{table1} Age-optimal update selection for fixed $k=5$ with different arrival rates, $\lambda$.}
  \vspace{-5 mm}
\end{table*}

Thus, even though the highest $k$ selective encoding policy improves the age performance as shown in Section~\ref{sect:num_res}, this selection may not necessarily be optimal for a given pmf and arrival rate among all other possible selections. In fact, in Table~\ref{table1} we observe that, the highest $k$ selection is optimal when the arrival rate is low. When the arrival rate is high, however, a different $k$ selection should be implemented to get a better age performance as shown in Table~\ref{table1}. The theoretical analysis for the optimality of the highest $k$ selective encoding remains as a future work. Further, in some cases the realizations with lower probabilities may carry important information that cannot be ignored. In these scenarios, an importance metric can be assigned to each realization and the encoded $k$ realizations can be selected considering both the importance metric and realization probabilities. We leave this problem as a future work.

\section{Conclusions} \label{sect:conc}
We considered a status updating system in which an information source generates independent and identically distributed update packets based on an observed random variable $X$ which takes $n$ values based on a known pmf. We studied three different encoding schemes for the transmitter node to send the realizations to the receiver node. In all these schemes, the most probable $k$ update realizations are always encoded. For the remaining less probable $n-k$ realizations, we considered the case in which these realizations are completely discarded, i.e., the highest $k$ selective encoding scheme. Next, we considered the case in which the remaining previously discarded $n-k$ realizations are encoded into codewords randomly to further inform the receiver, i.e., randomized selective encoding scheme. Lastly, we examined the case where the remaining less probable realizations are mapped into an empty symbol to partially inform the receiver node, i.e., highest $k$ selective encoding scheme with an empty symbol. We derived the average age for all these encoding schemes and determined the age-optimal codeword lengths. Through numerical results we showed that the proposed selective encoding scheme achieves a lower average age than encoding all the realizations, and determined the age-optimal $k$ values for arbitrary pmfs. 

\bibliographystyle{unsrt}
\bibliography{IEEEabrv,lib_v5}

\begin{thebibliography}{10}

\bibitem{MelihBatu1}
M.~Bastopcu, B.~Buyukates, and S.~Ulukus.
\newblock Optimal selective encoding for timely updates.
\newblock In {\em CISS}, March 2020.

\bibitem{Kaul12a}
S.~K. Kaul, R.~D. Yates, and M.~Gruteser.
\newblock Real-time status: How often should one update?
\newblock In {\em IEEE Infocom}, March 2012.

\bibitem{Costa14}
M.~Costa, M.~Codrenau, and A.~Ephremides.
\newblock Age of information with packet management.
\newblock In {\em IEEE ISIT}, June 2014.

\bibitem{Bedewy16}
A.~M. Bedewy, Y.~Sun, and N.~B. Shroff.
\newblock Optimizing data freshness, throughput, and delay in multi-server
  information-update systems.
\newblock In {\em IEEE ISIT}, July 2016.

\bibitem{He16a}
Q.~He, D.~Yuan, and A.~Ephremides.
\newblock Optimizing freshness of information: On minimum age link scheduling
  in wireless systems.
\newblock In {\em IEEE WiOpt}, May 2016.

\bibitem{Kam16b}
C.~Kam, S.~Kompella, G.~D. Nguyen, Wieselthier~J. E., and A.~Ephremides.
\newblock Age of information with a packet deadline.
\newblock In {\em IEEE ISIT}, July 2016.

\bibitem{Sun17a}
Y.~Sun, E.~Uysal-Biyikoglu, R.~D. Yates, C.~E. Koksal, and N.~B. Shroff.
\newblock Update or wait: How to keep your data fresh.
\newblock {\em IEEE Transactions on Information Theory}, 63(11):7492--7508,
  November 2017.

\bibitem{Najm18b}
E.~Najm and E.~Telatar.
\newblock Status updates in a multi-stream {M/G/1/1} preemptive queue.
\newblock In {\em IEEE Infocom}, April 2018.

\bibitem{Najm17}
E.~Najm, R.~D. Yates, and E.~Soljanin.
\newblock Status updates through {M/G/1/1} queues with {HARQ}.
\newblock In {\em IEEE ISIT}, June 2017.

\bibitem{Soysal18}
A.~Soysal and S.~Ulukus.
\newblock Age of information in {G/G/1/1} systems.
\newblock In {\em Asilomar Conference}, November 2019.

\bibitem{Soysal19}
A.~Soysal and S.~Ulukus.
\newblock Age of information in {G/G/1/1} systems: Age expressions, bounds,
  special cases, and optimization.
\newblock May 2019.
\newblock Available on arXiv: 1905.13743.

\bibitem{Yates17b}
R.~D. Yates, P.~Ciblat, A.~Yener, and M.~Wigger.
\newblock Age-optimal constrained cache updating.
\newblock In {\em IEEE ISIT}, June 2017.

\bibitem{Tang19}
H.~Tang, P.~Ciblat, J.~Wang, M.~Wigger, and R.~D. Yates.
\newblock Age of information aware cache updating with file- and age-dependent
  update durations.
\newblock September 2019.
\newblock Available on arXiv: 1909.05930.

\bibitem{Nath17}
S.~Nath, J.~Wu, and J.~Yang.
\newblock Optimizing age-of-information and energy efficiency tradeoff for
  mobile pushing notifications.
\newblock In {\em IEEE SPAWC}, July 2017.

\bibitem{Hsu18b}
Y.~Hsu.
\newblock Age of information: Whittle index for scheduling stochastic arrivals.
\newblock In {\em IEEE ISIT}, June 2018.

\bibitem{Kadota18a}
I.~Kadota, A.~Sinha, E.~Uysal-Biyikoglu, R.~Singh, and E.~Modiano.
\newblock Scheduling policies for minimizing age of information in broadcast
  wireless networks.
\newblock {\em IEEE/ACM Transactions on Networking}, 26(6):2637--2650, December
  2018.

\bibitem{Buyukates18c}
B.~Buyukates, A.~Soysal, and S.~Ulukus.
\newblock Age of information scaling in large networks.
\newblock In {\em IEEE ICC}, May 2019.

\bibitem{Buyukates19b}
B.~Buyukates, A.~Soysal, and S.~Ulukus.
\newblock Age of information scaling in large networks with hierarchical
  cooperation.
\newblock In {\em IEEE Globecom}, December 2019.

\bibitem{Gong19}
J.~Gong, Q.~Kuang, X.~Chen, and X.~Ma.
\newblock Reducing age-of-information for computation-intensive messages via
  packet replacement.
\newblock January 2019.
\newblock Available on arXiv: 1901.04654.

\bibitem{Buyukates19c}
B.~Buyukates and S.~Ulukus.
\newblock Timely distributed computation with stragglers.
\newblock October 2019.
\newblock Available on arXiv: 1910.03564.

\bibitem{Arafa19b}
A.~Arafa, K.~Banawan, K.~G. Seddik, and H.~V. Poor.
\newblock On timely channel coding with hybrid {ARQ}.
\newblock In {\em IEEE Globecom}, December 2019.

\bibitem{Sun17b}
Y.~Sun, Y.~Polyanskiy, and E.~Uysal-Biyikoglu.
\newblock Remote estimation of the {Wiener} process over a channel with random
  delay.
\newblock In {\em IEEE ISIT}, June 2017.

\bibitem{Sun18b}
Y.~Sun and B.~Cyr.
\newblock Information aging through queues: A mutual information perspective.
\newblock In {\em IEEE SPAWC}, June 2018.

\bibitem{Chakravorty18}
J.~Chakravorty and A.~Mahajan.
\newblock Remote estimation over a packet-drop channel with {Markovian} state.
\newblock July 2018.
\newblock Available on arXiv:1807.09706.

\bibitem{Bastopcu19}
M.~Bastopcu and S.~Ulukus.
\newblock Age of information for updates with distortion.
\newblock In {\em IEEE ITW}, August 2019.

\bibitem{Bastopcu20b}
M.~Bastopcu and S.~Ulukus.
\newblock Age of information for updates with distortion: Constant and
  age-dependent distortion constraints.
\newblock December 2019.
\newblock Available on arXiv:1912.13493.

\bibitem{Bastopcu20c}
M.~Bastopcu and S.~Ulukus.
\newblock Who should {Google} {Scholar} update more often?
\newblock In {\em Infocom Workshop on Age of Information}, July 2020.

\bibitem{partial_updates}
D.~Ramirez, E.~Erkip, and H.~V. Poor.
\newblock Age of information with finite horizon and partial updates.
\newblock October 2019.
\newblock Available on arXiv:1910.00963.

\bibitem{Zou19b}
P.~Zou, O.~Ozel, and S.~Subramaniam.
\newblock Trading off computation with transmission in status update systems.
\newblock In {\em IEEE PIMRC}, September 2019.

\bibitem{Non_linear}
A.~Kosta, N.~Pappas, A.~Ephremides, and V.~Angelakis.
\newblock Age and value of information: Non-linear age case.
\newblock In {\em IEEE ISIT}, June 2017.

\bibitem{Bastopcu18}
M.~Bastopcu and S.~Ulukus.
\newblock Age of information with soft updates.
\newblock In {\em Allerton Conference}, October 2018.

\bibitem{bastopcu_soft_updates_journal}
M.~Bastopcu and S.~Ulukus.
\newblock Minimizing age of information with soft updates.
\newblock {\em Journal of Communications and Networks}, 21(3):233--243, June
  2019.

\bibitem{Arafa17b}
A.~Arafa and S.~Ulukus.
\newblock Age minimization in energy harvesting communications:
  Energy-controlled delays.
\newblock In {\em Asilomar Conference}, October 2017.

\bibitem{Arafa17a}
A.~Arafa and S.~Ulukus.
\newblock Age-minimal transmission in energy harvesting two-hop networks.
\newblock In {\em IEEE Globecom}, December 2017.

\bibitem{Wu18}
X.~Wu, J.~Yang, and J.~Wu.
\newblock Optimal status update for age of information minimization with an
  energy harvesting source.
\newblock {\em IEEE Transactions on Green Communications and Networking},
  2(1):193--204, March 2018.

\bibitem{Arafa_Age_Online}
A.~Arafa, J.~Yang, and S.~Ulukus.
\newblock Age-minimal online policies for energy harvesting sensors with random
  battery recharges.
\newblock In {\em IEEE ICC}, May 2018.

\bibitem{Arafa18a}
A.~Arafa, J.~Yang, S.~Ulukus, and H.~V. Poor.
\newblock Age-minimal online policies for energy harvesting sensors with
  incremental battery recharges.
\newblock In {\em UCSD ITA}, February 2018.

\bibitem{Arafa18f}
A.~Arafa, J.~Yang, S.~Ulukus, and H.~V. Poor.
\newblock Online timely status updates with erasures for energy harvesting
  sensors.
\newblock In {\em Allerton Conference}, October 2018.

\bibitem{Arafa19e}
A.~Arafa, J.~Yang, S.~Ulukus, and H.~V. Poor.
\newblock Using erasure feedback for online timely updating with an energy
  harvesting sensor.
\newblock In {\em IEEE ISIT}, July 2019.

\bibitem{Farazi18}
S.~Farazi, A.~G. Klein, and D.~R. Brown~III.
\newblock Average age of information for status update systems with an energy
  harvesting server.
\newblock In {\em IEEE Infocom}, April 2018.

\bibitem{Yener_energy_19}
S.~Leng and A.~Yener.
\newblock Age of information minimization for an energy harvesting cognitive
  radio.
\newblock {\em IEEE Transactions on Cognitive Communications and Networking},
  5(2):427--439, May 2019.

\bibitem{Chen19}
Z.~Chen, N.~Pappas, E.~Bjornson, and E.~G. Larsson.
\newblock Age of information in a multiple access channel with heterogeneous
  traffic and an energy harvesting node.
\newblock March 2019.
\newblock Available on arXiv: 1903.05066.

\bibitem{Elmagid18}
M.~A. Abd-Elmagid and H.~S. Dhillon.
\newblock Average peak age-of-information minimization in {UAV}-assisted {IoT}
  networks.
\newblock {\em IEEE Transactions on Vehicular Technology}, 68(2):2003--2008,
  February 2019.

\bibitem{Liu18}
J.~Liu, X.~Wang, and H.~Dai.
\newblock Age-optimal trajectory planning for {UAV}-assisted data collection.
\newblock In {\em IEEE Infocom}, April 2018.

\bibitem{Ceran18}
E.~T. Ceran, D.~Gunduz, and A.~Gyorgy.
\newblock A reinforcement learning approach to age of information in multi-user
  networks.
\newblock In {\em IEEE PIMRC}, September 2018.

\bibitem{Beytur19}
H.~B. Beytur and E.~Uysal-Biyikoglu.
\newblock Age minimization of multiple flows using reinforcement learning.
\newblock In {\em IEEE ICNC}, February 2019.

\bibitem{Elmagid19}
M.~A. Abd-Elmagid, H.~S. Dhillon, and N.~Pappas.
\newblock A reinforcement learning framework for optimizing age-of-information
  in {RF}-powered communication systems.
\newblock August 2019.
\newblock Available on arXiv: 1908.06367.

\bibitem{Mayekar18}
P.~Mayekar, P.~Parag, and H.~Tyagi.
\newblock Optimal lossless source codes for timely updates.
\newblock In {\em IEEE ISIT}, June 2018.

\bibitem{Zhong16}
J.~Zhong and R.~D. Yates.
\newblock Timeliness in lossless block coding.
\newblock In {\em IEEE DCC}, March 2016.

\bibitem{Yates_Soljanin_source_coding}
J.~Zhong, R.~D. Yates, and E.~Soljanin.
\newblock Timely lossless source coding for randomly arriving symbols.
\newblock In {\em IEEE ITW}, November 2018.

\bibitem{Zhong17a}
J.~Zhong, E.~Soljanin, and R.~D. Yates.
\newblock Status updates through multicast networks.
\newblock In {\em Allerton Conference}, October 2017.

\bibitem{Zhong18b}
J.~Zhong, R.~D. Yates, and E.~Soljanin.
\newblock Multicast with prioritized delivery: How fresh is your data?
\newblock In {\em IEEE SPAWC}, June 2018.

\bibitem{Buyukates18}
B.~Buyukates, A.~Soysal, and S.~Ulukus.
\newblock Age of information in two-hop multicast networks.
\newblock In {\em Asilomar Conference}, October 2018.

\bibitem{Buyukates18b}
B.~Buyukates, A.~Soysal, and S.~Ulukus.
\newblock Age of information in multihop multicast networks.
\newblock {\em Journal of Communications and Networks}, 21(3):256--267, July
  2019.

\bibitem{Buyukates19}
B.~Buyukates, A.~Soysal, and S.~Ulukus.
\newblock Age of information in multicast networks with multiple update
  streams.
\newblock In {\em Asilomar Conference}, November 2019.

\bibitem{Cover}
T.~M. Cover and J.~A. Thomas.
\newblock {\em Elements of Information Theory}.
\newblock Wiley Press, 2012.

\bibitem{Yates14}
R.~D. Yates and D.~J. Goodman.
\newblock {\em Probability and Stochastic Processes}.
\newblock Wiley, 2014.

\bibitem{frac_programming}
W.~Dinkelbach.
\newblock On nonlinear fractional programming.
\newblock {\em Management Science}, 13(7):435--607, March 1967.

\bibitem{Boyd04}
S.~P. Boyd and L.~Vandenberghe.
\newblock {\em Convex Optimization}.
\newblock Cambridge University Press, 2004.

\bibitem{lambert}
R.~M. Corless, G.~H. Gonnet, D.~E.~G. Hare, D.~J. Jeffrey, and D.~E. Knuth.
\newblock On the {Lambert} {W} function.
\newblock {\em Advances in Computational Mathematics}, 5(1):329--359, December
  1996.

\bibitem{Bastopcu20a}
M.~Bastopcu and S.~Ulukus.
\newblock Partial updates: Losing information for freshness.
\newblock January 2020.
\newblock Available on arXiv: 2001.11014.

\end{thebibliography}
\end{document}